\documentclass[10pt, conference, compsocconf]{IEEEtran}
\usepackage{algorithm,algorithmic,amsbsy,amsmath,amssymb,epsfig,bbm,mathrsfs,multirow,amsthm}
\usepackage{subfigure,float,url}

\usepackage[top=0.75in, bottom=0.67in, left=0.67in, right=0.67in]{geometry}

\begin{document}
\title{A Stochastic Programming Approach for Risk Management in Mobile Cloud Computing}
\author{Dinh Thai Hoang$^1$, Dusit Niyato$^1$, Ping Wang$^1$, Shaun Shuxun Wang$^2$, Diep Nguyen$^3$, and Eryk Dutkiewicz$^3$ \\
$^1$ School of Computer Science and Engineering, Nanyang Technological University, Singapore \\
$^2$ Nanyang Business School, Nanyang Technological University, Singapore \\
$^3$ School of Computing and Communications, University of Technology, Sydney, Australia	\vspace{-5mm}}

\maketitle
%++++++++++++++++++++++++++++++++++++++++++++++++++++++++++++++++++++
%++++++++++++++++++++++++++++++++++++++++++++++++++++++++++++++++++++
\begin{abstract}
The development of mobile cloud computing has brought many benefits to mobile users as well as cloud service providers. However, mobile cloud computing is facing some challenges, especially security-related problems due to the growing number of cyberattacks which can cause serious losses. In this paper, we propose a dynamic framework together with advanced risk management strategies to minimize losses caused by cyberattacks to a cloud service provider. In particular, this framework allows the cloud service provider to select appropriate security solutions, e.g., security software/hardware implementation and insurance policies, to deal with different types of attacks. Furthermore, the stochastic programming approach is adopted to minimize the expected total loss for the cloud service provider under its financial capability and uncertainty of attacks and their potential losses. Through numerical evaluation, we show that our approach is an effective tool in not only dealing with cyberattacks under uncertainty, but also minimizing the total loss for the cloud service provider given its available budget. 
\end{abstract}
{\it Keywords-} Cyber insurance, cybersecurity, cloud services, mobile cloud, stochastic programming. 
 
%++++++++++++++++++++++++++++++++++++++++++++++++++++++++++++++++++++
%++++++++++++++++++++++++++++++++++++++++++++++++++++++++++++++++++++
\section{Introduction}

Mobile cloud computing (MCC) is an emerging platform using cloud computing to provide applications and services to mobile users through mobile networks. By taking advantages of cloud computing technology, mobile applications can be performed more efficiently, thereby generating huge profits for mobile users as well as cloud service providers (CSPs)~\cite{Hoang2013MCC}. Forbes magazine predicts that worldwide spending on public cloud services will grow at a 19.4\% compound annual growth rate (CAGR) from nearly \$70 billion in 2015 to more than \$141 billion in 2019. However, the development of MCCs is facing with some security-related challenges due to the growing of cyberattacks in the last few years. According to US government statistics, the number of ransomware attacks increased 300\% from 2015, with over 4,000 attacks detected per day in 2016~\cite{Ransomware2016}. Hence, countermeasures and risk management solutions for cybercrime are urgently in need.

Although there are many security solutions implemented to detect and prevent cyberattacks in MCC such as implementing firewall and installing antivirus software, achieving complete security protection is still nearly impossible~\cite{Pal2014Will}. Therefore, cyber insurance has been emerging as an alternative approach to address and manage cyber risks for cloud and computer networks~\cite{Marotta2017Cyber,Hoang2017Charging}. Under cyber insurance coverage, when an attack happens to a CSP, the CSP's losses will be covered partially or fully by the cyber insurance provider. In other words, cyber risks of the CSP now are ``transferred'' to the insurance provider through a cyber insurance contract. Thus, the CSP's risks can be mitigated significantly. However, cyber insurance is not always the best solution for the CSP because of the variety of attacks, the diversity of premiums and claims, and the CSP's limited budget. Hence, how to balance between security and insurance policies given a limited budget and under uncertainty of cyber threats is an important challenge.  

In this paper, we propose a dynamic framework which can utilize advantages of both security and insurance polices, and it can be applied widely in cloud environment to deal with cyber risks under uncertainty. This framework consists of two decision stages. In the first stage, the CSP has to decide how much it should invest to buy security packages to prevent cyberattacks and how much it should spend to buy insurance packages to cover losses caused by the attacks. The more money the CSP invests into security packages, the higher chance the cyberattacks can be successfully prevented, but the less money the CSP can spend for insurance. Then, in the second stage, the CSP observes actual attacks associated with their direct losses, and makes a decision to whether implement countermeasures to mitigate their indirect losses or not. To address the multi-stage optimization problem under uncertainty of attacks and a limited budget, we adopt the stochastic programming method~\cite{Wallace2005Applications} to find optimal budget allocation policies for the CSP. Through performance evaluation, we demonstrate that the proposed approach can use efficiently an available budget to minimize the total cost incurred by the cyberattacks under uncertainty. We also show the efficiency of using cyber insurance in dealing with cyber risks in MCC through balancing budget allocation. 

%The rest of the paper is organized as follows. Section~\ref{sec:RW} presents related works. In Section~\ref{sec:SM} and Section~\ref{sec:PS}, we introduce the proposed system model, time stages, and decision variables. Then, the stochastic optimization problem is formulated in Section~\ref{sec:PF}, and the performance of the proposed solution is evaluated in Section~\ref{sec:PE}. Finally, conclusions and future research works are presented in Section~\ref{sec:Sum}. 

%++++++++++++++++++++++++++++++++++++++++++++++++++++++++++++++++++++
%++++++++++++++++++++++++++++++++++++++++++++++++++++++++++++++++++++
\section{Related Work}
\label{sec:RW}

The rapid growth of cloud services together with high security demands recently have brought a great opportunity for cyber insurance providers. The first commercial cyber insurance products were introduced by \emph{Cloudinsurnce} (http://www.cloudinsure.com) in 2013 to specifically address emerging liabilities within the cloud environment. In partnership with global insurance and reinsurance carriers, Cloudinsurnce developed privacy and security liability coverage to meet security demands of cloud service providers (CSPs). Recently, \emph{MSPAlliance} (http://www.mspalliance.com) in partnership with Lockton Affinity, the world's largest privately owned independent insurance broker, announced a new cloud insurance service, called, Cloud and MSP Insurance~\cite{CloudMSPInsurance}. This insurance may be included as part of a service level agreement with the CSP. Alternatively, it may be purchased separately through a third-party insurance company which works with the CSP. While Cloudinsurance and MSPAlliance aim to provide insurance policies for CSPs, \emph{Cloudsurance} (http://www.cloudsurance.com) released in 2016 is the first cyber cloud insurance package designed to offer financial protection to cloud consumers. Cloudsurance offers a wide range of cyber insurance services to cloud customers such as data loss compensation, cyberattack protection, and urgent migration support. The emergence of Cloudsurance has made the cyber insurance market more and more diverse and promising. 

In the literature, applications of cyber insurance in the cloud environment have also received a lot of attention. In particular, the authors in~\cite{Elnagdy2016Understanding} presented a taxonamy of cyber risks and discussed risk management methods for cybersecurity insurance in cloud computing. Findings in this paper can be used as a reference for further research in cybersecurity insurance. In~\cite{Gai2016Anovel}, the authors  proposed a secure cyber incident framework to reduce cyber insurance costs without lowering down the security level. To do so, the authors introduced an optimal cost balance algorithm to achieve the highest cybersecurity guarantee given a financial budget constraint. The numerical results showed that cyber insurance is an effective way to mitigate losses caused by cyberattacks for cloud providers. Alternatively, there are some other work introducing applications of cyber insurance to deal with cyber risks for cloud customers~\cite{Chaisiri2015Ajoint, Chase2017Ascalable}. The core idea of these papers is using stochastic programming to balance provisioning of security and insurance, even when future costs and risks are uncertain. The authors demonstrated the effectiveness of the proposed solution through evaluation on real attack data. 

Different from all aforementioned research works and others in the literature, in this paper, we introduce a stochastic programming framework to minimize the total cost for the CSP. This framework not only utilizes advantages of security and insurance policies, but also allows the CSP to implement prompt countermeasures to mitigate indirect losses of cyberattacks under uncertainty and a limited budget.

%\clearpage
%++++++++++++++++++++++++++++++++++++++++++++++++++++++++++++++++++++
%++++++++++++++++++++++++++++++++++++++++++++++++++++++++++++++++++++
\section{System Model}
\label{sec:SM}

\begin{figure}[h]
	\begin{center}
		\epsfxsize=3.5 in \epsffile{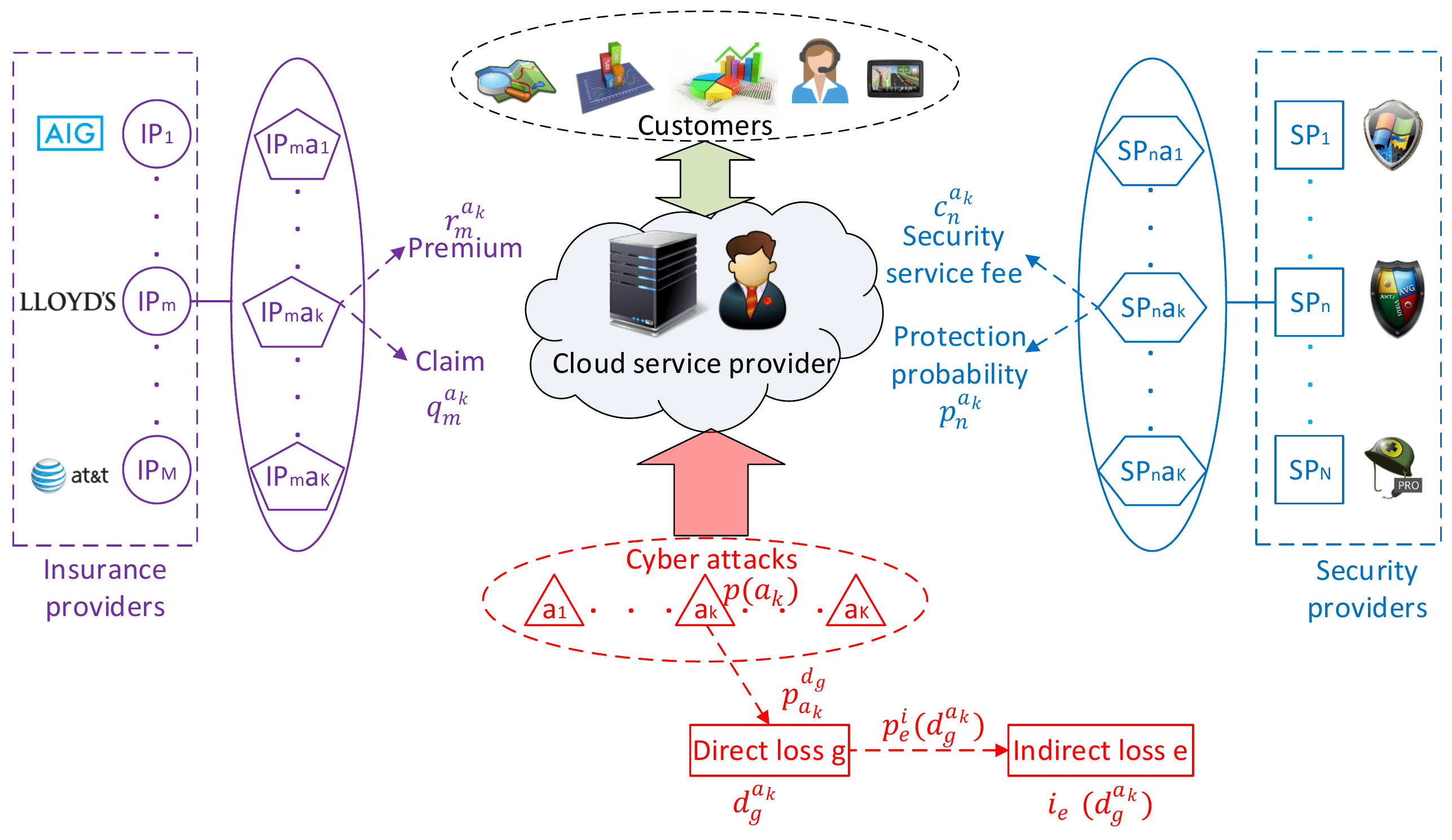} 
		\caption{System model.}
		\label{fig:SystemModel}
	\end{center}
\end{figure}

The system model is illustrated in Fig.~\ref{fig:SystemModel}. In this model, the CSP provides cloud services/applications to its customers. When the CSP provides services, it can suffer from cyberattacks which can cause serious damages. There are $K$ types of attacks which can occur with different probabilities in the business time. We denote $\mathcal{A} = \{a_1,\ldots,a_k,\ldots,a_K\}$ as the set of attacks which can occur simultaneously at the business time with probabilities $\{p(a_1),\ldots,p(a_k),\ldots,p(a_K)\}$, respectively. Therefore, before the service period, the CSP can choose to buy security packages to prevent cyberattacks which can occur at the business time. 

We assume that there are $N$ security service providers (SPs), and each SP provides $K$ security packages to deal with $K$ types of attacks. The security package offered by SP-$n$ to prevent attack $a_k$ is denoted by $s_n^{a_k}$. Each $s_n^{a_k}$ costs $c_n^{a_k}$, and it can protect the CSP from attack $a_k$ with probability $p_{n}^{a_k}$. Given the available budget and information of attacks (e.g., probabilities of attacks and their potential losses), the CSP can choose optimal security packages to minimize the potential loss. However, when attacks happen, the implemented security solutions may not be able to prevent the attacks completely. Thus, the CSP can choose to buy cyber insurance packages to cover its services when cyber risks happen. 

We assume that there are $M$ cyber insurance providers (IPs), and each IP offers $K$ insurance packages for $K$ types of attacks. The insurance package offered by IP-$m$ to cover the loss caused by attack $a_k$ is denoted by $i_m^{a_k}$. Each $i_m^{a_k}$ has premium $r_m^{a_k}$ and covers $q_m^{a_k} \%$ of the loss of attack $a_k$. Here, there are two types of losses, i.e., direct loss and indirect loss, with the amounts denoted by $d^{a_k}_g$ and $i_e(d^{a_k}_g)$, respectively. Direct loss refers to a damage immediately inflicted by an attack (e.g., business interruption), while indirect loss refers to an indirect effect of an attack (e.g., damage to company's reputation). For example, when a denial-of-service (DoS) attack happens, customers cannot access cloud services, and this causes direct loss to the CSP because there is no revenue from the customers and the CSP must repair the system. If the repair time is too long, the CSP's reputation will be reduced, and thus the CSP may lose the customers in the future. 

Currently, cyber insurance packages offered by IPs, e.g., AIG (http://www.AIG.com) and Cloudsurance, cover only direct losses such as data loss and downtime compensation. Thus, in this paper, insurance packages of IPs are assumed to cover only direct losses from cyberattacks. To mitigate indirect losses, the CSP can adopt repair packages to make the system back to work as soon as possible, thereby reducing indirect losses of cyberattacks. These repair packages can be provided by a third-party or by the CSP itself. We assume that there are $U$ repair packages from which the CSP can choose, and each repair package $u \in \mathcal{U} = \{1,\ldots,U\}$ is associated with a repair package fee $s_u^{d_g,a_k}$ and a percentage of damage reduction if we implement repair package $u$. We denote $1-t_u^{d_g,a_k}$ as the percentage of indirect losses which can be reduced if we implement repair package $u$ for direct loss $d_g$ under attack $a_k$. The repair packages are especially useful in dealing with cyberattacks with high indirect losses. However, due to the limited budget and the diverse attacks and their losses, the CSP needs to balance expenditure distribution to minimize its total cost.

%\clearpage
%++++++++++++++++++++++++++++++++++++++++++++++++++++++++++++++++++++
%++++++++++++++++++++++++++++++++++++++++++++++++++++++++++++++++++++
\section{Stages, Decision Variables, and Uncertainty}
\label{sec:PS}

\begin{figure}[!th]
	\begin{center}
		\epsfxsize=3.4 in \epsffile{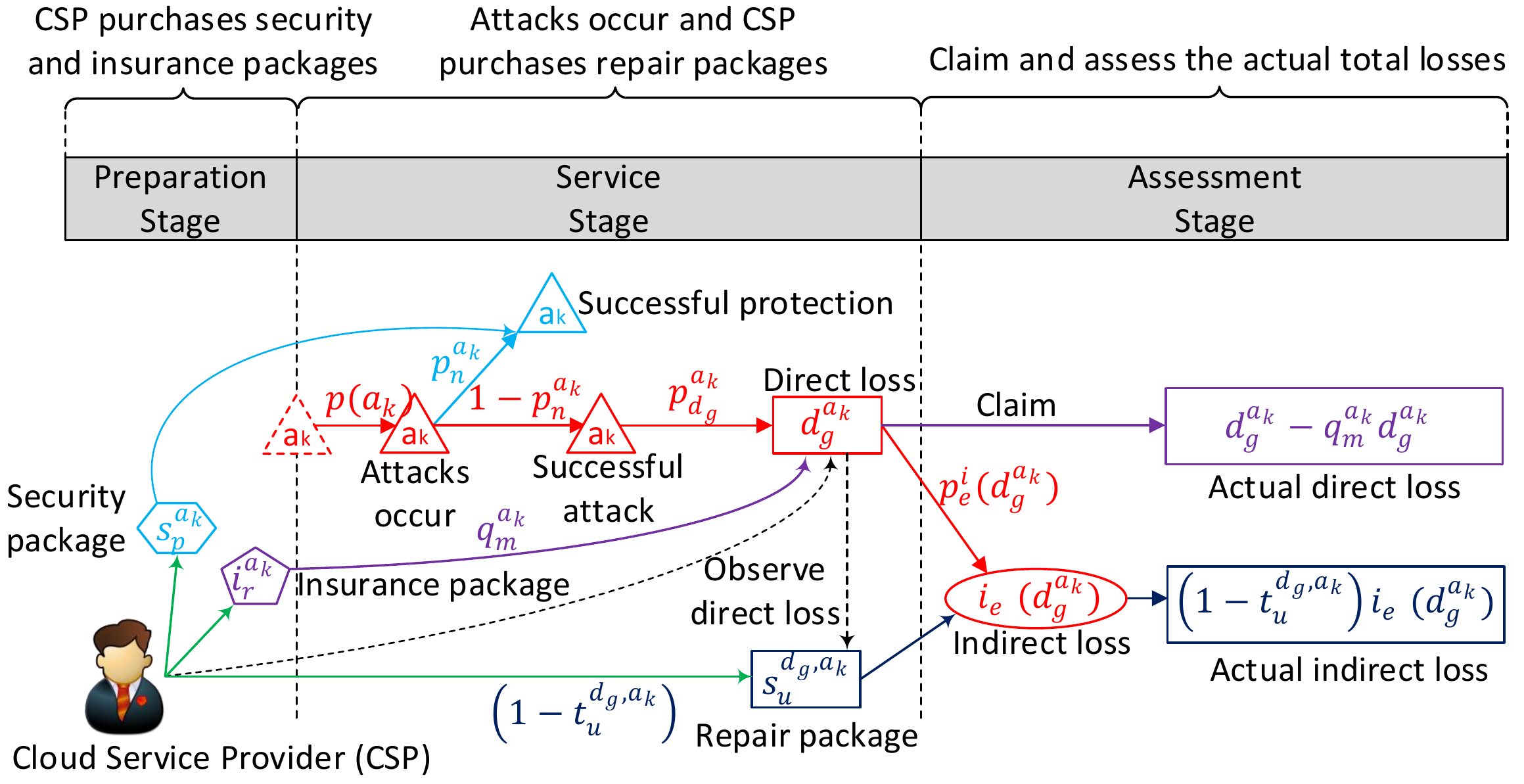} 
		\caption{In each business time period, there are three stages.}
		\label{fig:Stages}
	\end{center}
\end{figure}

The stochastic programming model under considerations contains three stages, i.e., preparation, service, and assessment stages, as illustrated in Fig.~\ref{fig:Stages}. At the first stage, the CSP evaluates potential losses of cyberattacks which can happen in the service stage, and makes decisions to buy security and insurance packages to prevent cyberattacks and mitigate their potential losses. Then, in the second stage, the CSP observes the actual attacks and their direct losses, and selects appropriate repair packages to prevent potential indirect losses. Finally, at the last stage, the CSP assesses the actual total loss. 

%===============================================
%===============================================
\subsection{Decision Variables} 

The solution of a stochastic optimization formulation is known as a decision, and is represented by a set of values assigned to the decision variables. In the considered system model, there are three decision variables which are made in the first two stages, i.e., preparation and service stages.

%-------------------------
%-------------------------
\subsubsection{Preparation stage}

In the preparation stage, the CSP has to make two decisions (1) which security packages and (2) which insurance packages should be purchased? If we denote $X_n^{a_k}$ as the decision variable to buy security package $n$ from $SP_n$ to prevent attack $a_k$ and $Y_m^{a_k}$ as the decision variable to buy insurance package $m$ from $IP_m$ to cover attack $a_k$, then we have:
\begin{equation}
\begin{aligned}
& X_n^{a_k} \in \{0,1\}	, \forall a_k \in \mathcal{A} \phantom{5} \text{and} \phantom{5} \forall n \in \mathcal{N} = \{1,\ldots,N\} ,	\\
& Y_m^{a_k} \in \{0,1\} , \forall a_k \in \mathcal{A} \phantom{5} \text{and} \phantom{5} \forall m \in \mathcal{M} = \{1,\ldots,M\}	,
\end{aligned}
\end{equation}
where $1$ is to express that the CSP agrees to buy, and $0$ otherwise. Here, variables $X_n^{a_k}$ and $Y_m^{a_k} $ must satisfy the following constraints:
\begin{equation}
\begin{aligned}
& X_n^{a_k} 	\in \{0,1\}, \phantom{5}	\forall a_k \in \mathcal{A}  \phantom{5} \text{and} \phantom{5} n \in \mathcal{N} ,  \\
& \text{and} \phantom{5} 	\sum_{n=1}^{N} X_n^{a_k} \leq 1, \phantom{5} \forall a_k \in \mathcal{A} 	.		
\label{cons_X}
\end{aligned}
\end{equation}
\begin{equation}
\begin{aligned}
& Y_m^{a_k} 	\in \{0,1\}, \phantom{5}	\forall a_k \in \mathcal{A} \phantom{5} \text{and} \phantom{5} m \in \mathcal{M} ,  \\
& \text{and} \phantom{5} 	\sum_{m=1}^{M} Y_m^{a_k} \leq 1, \phantom{5} \forall a_k \in \mathcal{A} 	.		
\label{cons_Y}
\end{aligned}
\end{equation}
\begin{equation}
\sum_{k=1}^{K} \sum_{n=1}^{N} X_n^{a_k} c_n^{a_k} + \sum_{k=1}^{K} \sum_{m=1}^{M} Y_m^{a_k} r_m^{a_k} 	\leq 	\mathscr{B}^*	.
\label{cons_X+Y}
\end{equation}
The constraints in~(\ref{cons_X}) and~(\ref{cons_Y}) are to ensure that each attack is covered by at most one security provider and one insurance provider, respectively. The constraint in~(\ref{cons_X+Y}) is to make sure that the total expenditure of the CSP does not exceed the available budget $\mathscr{B}^*$.

%-------------------------
%-------------------------
\subsubsection{Service stage}

When an attack $a_k$ happens and is successful, it can cause a direct loss $d_g^{a_k}$ with probability $p^{d_g}_{a_k}$. Here, we denote $G$ as the total number of direct loss cases, then we have:
\begin{equation}
\sum_{g=1}^{G} p^{d_g}_{a_k} = 1, \forall a_k \in \mathcal{A} .
\end{equation}

When direct loss $d_g^{a_k}$ happens, it will associate with indirect loss $i_e(d_g^{a_k})$ with probability $p_e^i(d_g^{a_k})$. We denote $E$ as the total number of indirect loss cases, then we have:
\begin{equation}
\sum_{e=1}^{E} p_e^i(d_g^{a_k}) = 1, \forall a_k \in \mathcal{A} \phantom{5} \text{and} \phantom{5} g \in \mathcal{G} = \{1,\ldots,G\} .
\end{equation}

If the direct loss $d_g^{a_k}$ happens, the CSP can choose a repair package to mitigate the indirect losses caused by the direct loss $d_g^{a_k}$. We denote $Z_u^{d_g,a_k}$ as the decision variable to implement repair package $u$ for direct loss $d_g$ under attack $a_k$. Then, we have the following constraints:
\begin{equation}
\begin{aligned}
& Z_u^{d_g,a_k} \in \{0,1\}, \phantom{5} \forall a_k \in \mathcal{A} ,  u \in \mathcal{U} \phantom{5} \text{and} \phantom{5} g \in \mathcal{G} ,  \\ 
& \text{and} \phantom{5} 	\sum_{u=1}^{U} Z_u^{d_g,a_k} \leq 1, \phantom{5} \forall a_k \in \mathcal{A} \phantom{5} \text{and} \phantom{5} \forall g \in \mathcal{G}	.
\label{cons_Z}
\end{aligned}
\end{equation}
\begin{equation}
\begin{aligned}
& \sum_{k=1}^{K} \sum_{n=1}^{N} X_n^{a_k} c_n^{a_k} 	+ \sum_{k=1}^{K} \sum_{m=1}^{M} Y_m^{a_k} r_m^{a_k}  \\
& + \sum_{k=1}^{K} \sum_{u=1}^{U} \sum_{g=1}^{G} Z_u^{d_g,a_k} s_u^{d_g,a_k} 	\leq 	\mathscr{B}^*	.
\label{cons_X+Y+Z}
\end{aligned}
\end{equation}

Here, the constraint in~(\ref{cons_Z}) is to ensure that each direct loss caused by an attack is covered by at most one repair package, and the constraint in~(\ref{cons_X+Y+Z}) is to make sure that the total expenditure of the CSP does not exceed the available budget $\mathscr{B}^*$.

%-------------------------
%-------------------------
\subsubsection{Assessment stage}

In the last stage, the CSP evaluates the actual indirect losses and makes claims for direct losses covered by the IPs.

%===============================================
%===============================================
\subsection{Uncertainty of Parameters and Scenarios} 

The considered system contains a number of uncertain parameters which are unknown in advance. In particular, the number of cyberattacks in the second stage, direct losses associated with each attack, and indirect losses of each attach given an actual direct loss, are unknown at the first stage. The number of attacks and their direct losses are only known in the second stage. Similarly, the indirect loss of each attack is only known in the third stage. We denote $\mathscr{A}$ and $\mathscr{D}$ as the set of all possible attack scenarios and the set of all possible direct loss scenarios associated with an attack, respectively. Let $\Omega^{\dagger}$ represent the set of all possible scenarios in the second stage, then $\Omega^{\dagger} = \mathscr{A} \times \mathscr{D}$, where $\times$ is the Cartesian product. Similarly, we denote $\mathscr{I}$ as the set of all possible indirect loss scenarios in the last stage. Then, the set of all possible scenarios in the last stage is $\Omega^{\ddagger} = \mathscr{I}$. 

%The terms $\pi_{\omega_1}$, $\pi_{\omega_2}$, and $\pi_{\omega_3}$, represent the corresponding probabilities of the scenarios. Scenario sets of each stage have probabilities summing to $1$.

%++++++++++++++++++++++++++++++++++++++++++++++++++++++++++++++++++++
%++++++++++++++++++++++++++++++++++++++++++++++++++++++++++++++++++++
\section{Problem Formulation}
\label{sec:PF}

%===============================================
%===============================================
\subsection{Stochastic Optimization Problem} 

To solve the problem, we adopt the stochastic programming optimization technique with multi-stage decisions. The optimization problem formulation is given in~(\ref{eq:obj_fun}).
%\begin{figure*}[!h]
%\normalsize
\begin{equation}
\begin{aligned}
\label{eq:obj_fun}
& \min_{X_n^{a_k}, Y_m^{a_k}, Z_u^{d_g,a_k}} \sum_{k=1}^{K}	\sum_{n=1}^{N} \sum_{m=1}^M \sum_{g=1}^G \sum_{u=1}^U	\sum_{e=1}^E	\Bigg( \mathscr{C}_{1} \big(X_n^{a_k}, Y_m^{a_k} \big) \\
& + \mathbf{E}_{\Omega^{\dagger}} \bigg[ \mathscr{C}_{2} \big( X_n^{a_k}, Y_m^{a_k}, Z_u^{d_g,a_k}, \omega^{\dagger} \big)	\\
& + \mathbf{E}_{\Omega^{\ddagger}} \Big[ \mathscr{C}_{3} \big( X_n^{a_k}, Y_m^{a_k}, Z_u^{d_g,a_k}, \omega^{\dagger}, \omega^{\ddagger} \big)		\Big] \bigg]	\Bigg) ,
\end{aligned}
\end{equation}
%\hrulefill
%\vspace*{-4pt}
%\end{figure*}
s.t.~(\ref{cons_X}),~(\ref{cons_Y}),~(\ref{cons_Z}), and~(\ref{cons_X+Y+Z}).

The objective function in~(\ref{eq:obj_fun}) aims to minimize the total cost of CSP in all three stages. The expected costs of the second and third stages are represented by $\mathbf{E}_{\Omega^{\dagger}}[\cdot]$ and $\mathbf{E}_{\Omega^{\ddagger}}[\cdot]$, respectively. Here, $\omega^{\dagger} \in \Omega^{\dagger}$ and $\omega^{\ddagger} \in \Omega^{\ddagger}$ are scenarios in the second and third stages, respectively.

In~(\ref{eq:obj_fun}), $\mathscr{C}_{1} \big(X_n^{a_k}, Y_m^{a_k} \big)$ is an optimization problem to minimize the total cost at the preparation stage, and it is defined as follows:
\begin{equation}
\begin{aligned}
& \mathscr{C}_{1} \big(X_n^{a_k}, Y_m^{a_k} \big)  =  \\
& \min_{X_n^{a_k}, Y_m^{a_k}} \Bigg( \sum_{k=1}^{K} \sum_{n=1}^{N} X_n^{a_k} c_n^{a_k} + \sum_{k=1}^{K} \sum_{m=1}^{M} Y_m^{a_k} r_m^{a_k} 	\Bigg)	.
\label{eq:obj_C1}
\end{aligned}
\end{equation}

Given fixed values of decision variables $X_n^{a_k}$ and a scenario in the second stage $\omega^{\dagger}$, we minimize the total cost in the second stage by:
\begin{equation}
\begin{aligned}
& \mathscr{C}_{2} \big(X_n^{a_k}, Y_m^{a_k}, Z_u^{d_g,a_k}, \omega^{\dagger} \big)  = \min_{Z_u^{a_k,d_g}}  \bigg( \sum_{k=1}^{K} \sum_{g=1}^{G}  d_g^{a_k} (X_n^{a_k}, \omega^{\dagger})  \\
& + \sum_{k=1}^{K} \sum_{u=1}^{U} \sum_{g=1}^{G} Z_u^{d_g,a_k} (X_n^{a_k}, \omega^{\dagger})	s_u^{d_g,a_k}  \bigg)	.
\label{eq:obj_C2}
\end{aligned}
\end{equation}

In~(\ref{eq:obj_C2}), $d_g^{a_k} (X_n^{a_k}, \omega^{\dagger})$ is the direct loss and $Z_u^{d_g,a_k} (X_n^{a_k}, \omega^{\dagger})$ is the  decision of the CSP given decisions $X_n^{a_k}$ ($\forall n \in \mathcal{N}$ and $a_k \in \mathcal{A}$) in the first stage and the scenario of attacks associated with their direct losses in the second stage. 

The total cost for the last stage is determined as follows:
\begin{equation}
\begin{aligned}
& \mathscr{C}_{3} \big( X_n^{a_k}, Y_m^{a_k}, Z_u^{d_g,a_k}, \omega^{\dagger}, \omega^{\ddagger}  \big)  =   \\
&  \sum_{e=1}^{E}  i_e ( X_n^{a_k}, Z_u^{d_g,a_k}, \omega^{\dagger}, \omega^{\ddagger}) - \sum_{k=1}^{K} \sum_{m=1}^{M}  Y_m^{a_k} ( \omega^{\dagger} ) q_m^{a_k}	.  
\label{eq:obj_C3}
\end{aligned}
\end{equation}

In~(\ref{eq:obj_C3}), the first and second terms represent the indirect loss and the claim of the CSP. Here, $i_e ( X_n^{a_k}, Z_u^{d_g,a_k}, \omega^{\dagger}, \omega^{\ddagger})$ is the indirect loss given decisions of the CSP made in the first and second stages and actual scenarios of the second and third stages. $Y_m^{a_k} ( \omega^{\dagger} ) q_m^{a_k} $ is the actual claim of the CSP given its decision to buy insurance in the first stage and the actual scenario in the second stage.

%===============================================
%===============================================
\subsection{Deterministic Equivalent Formulation} 

The aforementioned stochastic optimization formulation above can be transformed into a deterministic equivalent optimization formulation as follows~\cite{Kall1994Stochastic}:

%\begin{figure}[!h]
%\normalsize
\begin{equation}
\begin{aligned}
& \min_{X_n^{a_k}, Y_m^{a_k}, Z_u^{d_g,a_k}} \Big(	C1 	+	C2  + 	C3	\Big) , \\
& \text{s.t.}~(\ref{cons_X}),~(\ref{cons_Y}),~(\ref{cons_Z}), \text{and}~(\ref{cons_X+Y+Z}) 	,
\end{aligned}
\label{eq:Det_for}
\end{equation}
where
\begin{equation}
\begin{aligned}
& C1 = \sum_{k=1}^{K} \sum_{n=1}^{N} \Big[ X_n^{a_k} c_n^{a_k} \Big] + \sum_{k=1}^{K} \sum_{m=1}^{M} \Big[ Y_m^{a_k} r_m^{a_k} \Big]  ,
\end{aligned}
\end{equation}
\begin{equation}
\begin{aligned}
& C2 =  \sum_{k=1}^{K} \bigg[ p(a_k) \Big(1 - \sum_{n=1}^{N}  X_n^{a_k} p_n^{a_k} \Big)  \sum_{g=1}^{G}  p^{d_g}_{a_k} d_g^{a_k} \bigg]+ \\
& \sum_{k=1}^{K} \bigg[ p(a_k) \Big(1 - \sum_{n=1}^{N}  X_n^{a_k} p_n^{a_k} \Big) \sum_{g=1}^{G}  \Big( p^{d_g}_{a_k} \sum_{u=1}^{U} Z_u^{d_g,a_k} s_u^{d_g,a_k} \Big) \bigg] , 
\end{aligned}
\end{equation}
\begin{equation}
\label{eq:C3}
\begin{aligned}
& \text{and} \phantom{5} C3 = \sum_{k=1}^{K} \bigg[  p(a_k) \Big(1 - \sum_{n=1}^{N}  X_n^{a_k} p_n^{a_k} \Big) \\
& \sum_{g=1}^{G} \Big[ p^{d_g}_{a_k} \Big(1 - \sum_{u=1}^{U}  Z^{d_g,a_k}_u t^{d_g,a_k}_u \Big) \sum_{e=1}^{E} p_e^i(d_g^{a_k})  i_e(d_g^{a_k}) \Big]  \bigg] \\
& - \sum_{k=1}^{K} \bigg[ p(a_k) \Big(1 - \sum_{n=1}^{N}  X_n^{a_k} p_n^{a_k} \Big) \sum_{g=1}^{G}  p^{d_g}_{a_k} d_g^{a_k} \sum_{m=1}^{M}  Y_m^{a_k} q_m^{a_k}   \bigg] .
\end{aligned}
\end{equation}

Here, $C_1$, $C_2$, and $C_3$ represent the expected costs in the first, second, and third stages, respectively. In~(\ref{eq:C3}), the first and second terms represent the indirect loss and the claim of the CSP, respectively. If the CSP has purchased an insurance package to cover losses caused by attack $a_k$, the CSP can claim the direct loss caused by this attack, and thus the actual direct loss will be reduced by $q_m^{a_k} \%$. Similarly, if the CSP has bought the repair package to mitigate the indirect loss for attack $a_k$, the actual indirect loss will be reduced by $(1-t_u^g(a_k)) \%$.

%++++++++++++++++++++++++++++++++++++++++++++++++++++++++++++++++++++
%++++++++++++++++++++++++++++++++++++++++++++++++++++++++++++++++++++
\section{Performance Evaluation} 
\label{sec:PE}

%===============================================
%===============================================
\subsection{Experiment Setup} 

We consider two types of cyberattacks, e.g., \emph{data breaches} and \emph{DoS}, that are among the top 5 cyber threats in cloud environment~\cite{Top5Threats}. Data breaches are the most frequent attack in cloud computing and the number of data breach incidents accounts for around 40\% of the overall number of breaches in 2016~\cite{DataBreaches}. Losses caused by data breaches are diverse and they depend on many factors such as company size, the number of exposed records, and types of exposed data~\cite{MiniDataBreach}. Similarly, losses and probabilities of DoS attacks are diverse and they can be estimated based on some online support services, e.g.,~\cite{DDoSDowntimeCalculator}. In this paper, we consider a cloud online service provider, e.g., healthcare service provider, that is highly prone to cyber risk~\cite{CyberRiskHealthcare} with a medium size. In the simulation, we set the attack probability of data breaches at $p(a_2)=0.4$ and vary the DoS attack probability to evaluate the performance. For a medium-size service provider, the losses of DoS and data breaches can be set in the range from \$10,000 to \$100,000. Other parameters are provided in Table~\ref{tab:attacks}. Note that we normalize losses and fees in the range from \$10,000 to \$100,000 into $1$ to $10$ monetary units for presentation. 

\begin{table}[!]
	\centering % centering title
	\caption{Parameter Settings} 
	\centering % centering table
	\label{tab:attacks}
	\centering
	\begin{tabular}{|c||c|c||c|c||c|c||c|c||c|c||}
		\hline
		\textbf{Attacks} &\multicolumn{4}{c||}{$a_1$}&\multicolumn{4}{c||}{$a_2$}	\\
		%\cline{2-7}
		\hline
		$p(a_k)$ &\multicolumn{4}{c||}{$0.1$}&\multicolumn{4}{c||}{$0.4$} 	\\ 	\hline \hline
		\textbf{Direct loss}  &\multicolumn{2}{c||}{$d_1$}&\multicolumn{2}{c||}{$d_2$}&\multicolumn{2}{c||}{$d_1$}&\multicolumn{2}{c||}{$d_2$} 	\\ 	\hline
		$p^{d_g}_{a_k}$ &\multicolumn{2}{c||}{$0.3$}&\multicolumn{2}{c||}{$0.7$}&\multicolumn{2}{c||}{$0.2$}&\multicolumn{2}{c||}{$0.8$} 	\\ 	\hline	
		$d_g^{a_k}$ &\multicolumn{2}{c||}{$6.5$}&\multicolumn{2}{c||}{$5.3$}&\multicolumn{2}{c||}{$3.0$}&\multicolumn{2}{c||}{$4.0$} 	\\ 	\hline	 \hline
		\textbf{Indirect loss}  & $i_1$ & $i_2$ & $i_1$ & $i_2$ & $i_1$ & $i_2$ & $i_1$ & $i_2$	\\ \hline 
		$p^i_e(d_g^{a_k})$ & $0.2$ & $0.8$ & $0.6$ & $0.4$ & $0.6$ & $0.4$ & $0.5$ & $0.5$	\\ \hline	
		$i_e(d_g^{a_k})$ & $8.4$ & $3.4$ & $2.2$ & $5.1$ & $5.8$ & $5.4$ & $3.2$ & $2.8$ 	\\ \hline	\hline
		\textbf{Repair}  & $r_1$ & $r_2$ & $r_1$ & $r_2$ & $r_1$ & $r_2$ & $r_1$ & $r_2$	\\ \hline 
		$t_u^{d_g,a_k}$ & $0.5$ & $0.8$ & $0.5$ & $0.8$ & $0.5$ & $0.8$ & $0.5$ & $0.8$	\\ \hline	
		$s_u^{d_g,a_k}$ & $1.1$ & $1.9$ & $1.1$ & $1.9$ & $1.1$ & $1.9$ & $1.1$ & $1.9$ 	\\ \hline		
	\end{tabular}
\end{table}

\begin{table}[!]
	\centering % centering title
	\caption{The Security and Insurance Package Parameters} 
	\centering % centering table
	\label{tab:SP_IP}
	\centering
	\begin{tabular}{|c||c|c||c|c||c|c||c|c||}
		\hline
		&\multicolumn{2}{c||}{$SP_1$}&\multicolumn{2}{c||}{$SP_2$} &\multicolumn{2}{c||}{$IP_1$}&\multicolumn{2}{c||}{$IP_2$}	\\
		%\cline{2-7}
		\hline 
		& $c_n^{a_k}$ & $p_{n}^{a_k}$ & $c_n^{a_k}$ & $p_{n}^{a_k}$ & $r_m^{a_k}$ & $q_m^{a_k}\%$ & $r_m^{a_k}$ & $q_m^{a_k}\%$ 	\\ \hline
		$a_1$ & $1.2$ & $60\%$ & $1.8$ & $80\%$ & $0.6$ & $70\%$ & $0.9$ & $90\%$  	\\ \hline	
		$a_2$ & $1.4$ & $60\%$ & $1.9$ & $80\%$ & $0.8$ & $70\%$ & $1.1$ & $90\%$ 	\\ \hline		
	\end{tabular}
\end{table}

\begin{table*}[!t]
	\centering % centering title
	\caption{The Security Policies when the probability of attack $a_1$ is varied} 
	\centering % centering table
	\label{tab:Policy1}
	\centering
	\begin{tabular}{||c||c|c|c|c|c|c|c|c|c||}
		%\cline{2-7}
		\hline 
		Probability of attack $a_1$ & 0.1 & 0.2 & 0.3 & 0.4 & 0.5 & 0.6 & 0.7 & 0.8  & 0.9	\\ \hline \hline
		Security policy of attack $a_1$ & 0 & IP1 & IP2 & IP2 & IP2 & SP2 & SP2 & SP2 & SP2+IP1 	\\ \hline	
		$(a_1)$ Direct loss 1 & 0 & 0 & Rep1 & Rep1 & Rep1 & 0 & 0 & 0 & 0 	\\ \hline		
		$(a_1)$ Direct loss 2 & Rep1 & Rep1 & Rep1 & Rep1 & Rep1 & Rep1 & 0 &0 & 0 	\\ \hline	\hline
		Security policy of attack $a_2$ & IP1 & IP2 & SP2 & SP2 & SP2 & SP2 & IP1 & IP1 & SP2 	\\ \hline	
		$(a_2)$ Direct loss 1 & Rep2 & Rep1 & 0 & 0 & 0 & 0 & Rep1 & Rep1 & 0 	\\ \hline		
		$(a_2)$ Direct loss 2 & Rep1 & Rep1 & 0 & 0 & 0 & 0 & Rep1 & Rep1 & 0 	\\ \hline					
	\end{tabular}
\end{table*}

There are two security providers (SPs) and two insurance providers (IPs). $SP_1$ and $IP_1$ provide basic services, while $SP_2$ and $IP_2$ provide advanced services (i.e., higher fees with higher protection and claims) for the CSP. For example, the cost to implement Raptor Firewall NT v6.5 with virtual private network and unlimited mobile users is around \$18,000~\cite{FirewallCost}, while the implementation cost for the CoSoSys Endpoint Protector-4 to prevent data loss starts at \$8,250~\cite{CoSoSysProtector4}. Premiums for healthcare Software-as-a-Service providers are estimated around \$10,000~\cite{InsuranceCost}. Similar to Table~\ref{tab:attacks}, fees in Table~\ref{tab:SP_IP} are normalized in the range of $1$ to $10$. Other parameters of the SPs and IPs are given in Table~\ref{tab:SP_IP}. The budget for security policies of the CSP is limited at $5$ monetary units.

\begin{table*}[!t]
	\centering % centering title
	\caption{The Security Policies when the direct loss $d_1$ of attack $a_2$ is varied} 
	\centering % centering table
	\label{tab:Policy2}
	\centering
	\begin{tabular}{||c||c|c|c|c|c|c|c|c|c||}
		%\cline{2-7}
		\hline 
		Probability of direct loss $d_1$ under attack $a_2$ & 0.1 & 0.2 & 0.3 & 0.4 & 0.5 & 0.6 & 0.7 & 0.8  & 0.9	\\ \hline
		Security policy of attack $a_1$ & IP2 & IP2 & IP2 & IP2 & IP2 & IP2 & IP2 & IP2 & IP2 	\\ \hline	
		$(a_1)$ Direct loss 1 & Rep1 & Rep1 & Rep1 & Rep1 & 0 & 0 & Rep1 & Rep1 & Rep1 	\\ \hline		
		$(a_1)$ Direct loss 2 & Rep1 & Rep1 & Rep1 & Rep1 & Rep1 & Rep1 & Rep1 & Rep1 & Rep1 	\\ \hline	\hline
		Security policy of attack $a_2$ & IP1 & SP2 & SP2 & SP2 & IP1 & IP1 & 0 & 0 & 0 	\\ \hline	
		$(a_2)$ Direct loss 1 & 0 & 0 & 0 & 0 & Rep2 & Rep2 & Rep2 & Rep2 & Rep2 	\\ \hline		
		$(a_2)$ Direct loss 2 & Rep1 & 0 & 0 & 0 & 0 & 0 & 0 & 0 & 0 	\\ \hline		
	\end{tabular}
\end{table*}

%===============================================
%===============================================
\subsection{Numerical Results}

\begin{figure}[!]
	\begin{center}
		$\begin{array}{ccc} 
		\epsfxsize=1.65 in \epsffile{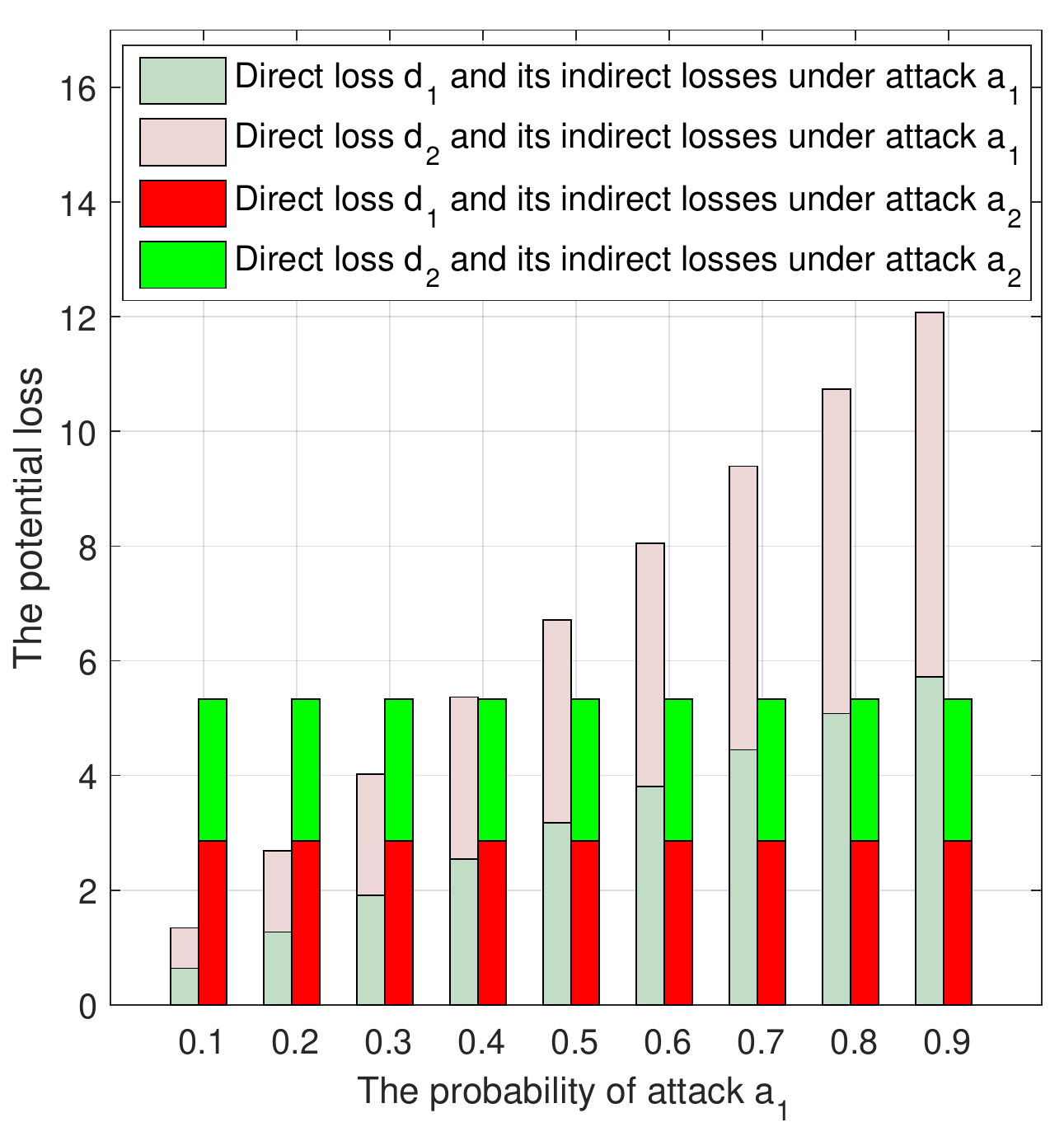} &
		\epsfxsize=1.65 in \epsffile{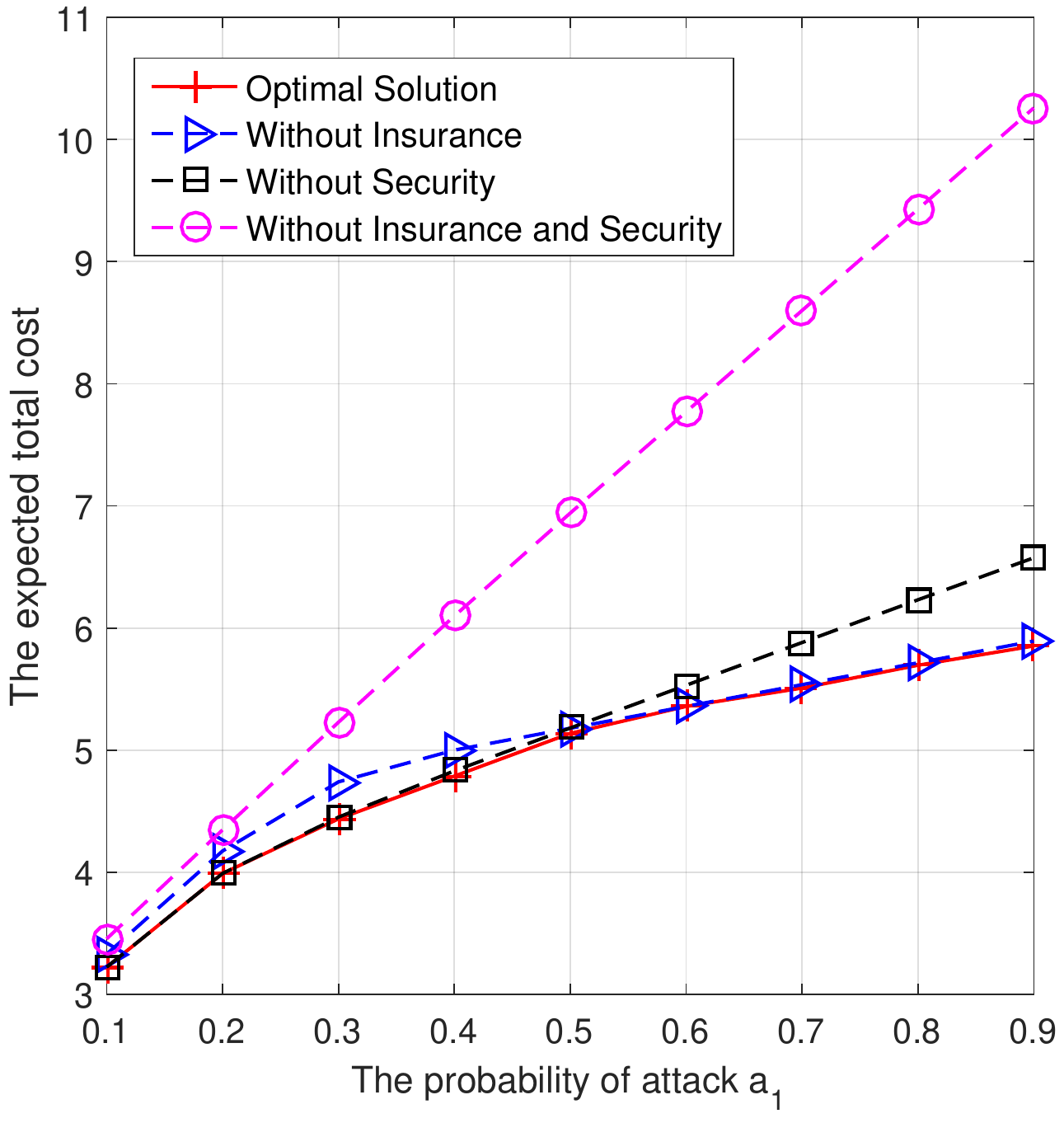} \\ [-0.2cm]
		(a) & (b) %& (c)
		\end{array}$
		\caption{(a) The expected costs of attacks and (b) the total expected cost when the probability of attack $a_1$ is varied.}
		\label{fig:vary_prob_a1}
	\end{center}
\end{figure}

We first vary the probability of attack $a_1$ and evaluate the performance of the proposed solution. In Table~\ref{tab:Policy1}, we show the security policy of the CSP to deal with the attacks, and Fig.~\ref{fig:vary_prob_a1} shows the performance of the proposed solution when the probability of attack $a_1$, i.e., $p(a_1)$, increases. As shown in Fig.~\ref{fig:vary_prob_a1}(a), when the probability of attack $a_1$ increases, the potential loss caused by attack $a_1$ (including its direct loss and indirect loss) will be increased. Thus, the CSP will invest more money to security solutions and insurance policies to mitigate the potential loss of attack $a_1$. However, the expenditure distribution is dissimilar under different $p(a_1)$. As observed in Table~\ref{tab:Policy1}, if the probability of attack $a_1$ is lower than $0.5$, the CSP will buy an insurance package together with a repair package to deal with attack $a_1$. Nevertheless, if the probability of attack $a_1$ is greater than $0.5$, the CSP will buy a security package to prevent attack $a_1$ at the first stage. The reason can be explained through Fig.~\ref{fig:vary_prob_a1}(a). When the probability of attack $a_1$ is lower than $0.5$, its occurrence probability and potential loss are not high, and thus the CSP can use insurance and repair policies to mitigate the potential loss of attack $a_1$. However, when the probability of attack $a_1$ is greater than $0.5$, both its occurrence probability and potential loss are very high. Hence, the CSP has to use a security package to prevent attack $a_1$ right at the first stage, thereby reducing both direct and indirect losses in the next stages. Note that since the CSP budget is limited at $5$, the CSP has to balance among security, insurance, and repair policies to minimize the expected total cost. This is an important issue especially for limited-budget companies in implementing countermeasures for cyberattacks.

In Fig.~\ref{fig:vary_prob_a1}(b), we compare the performance of the proposed solution in terms of the expected total cost with those of other approaches when the probability of attack $a_1$ is varied. Specifically, we compare with three other optimization approaches, i.e., optimization approaches without using insurance policies, without using security packages, and without using both security and insurance policies. As the probability of attack $a_1$ increases, the expected total costs of all approaches increase, and the cost obtained by the proposed solution is always the lowest one. The results in Fig.~\ref{fig:vary_prob_a1} demonstrate the importance in balancing between security and insurance policies to help the CSP to mitigate cyberattacks. 

\begin{figure}[!]
	\begin{center}
		$\begin{array}{cc} 
		\epsfxsize=1.60 in \epsffile{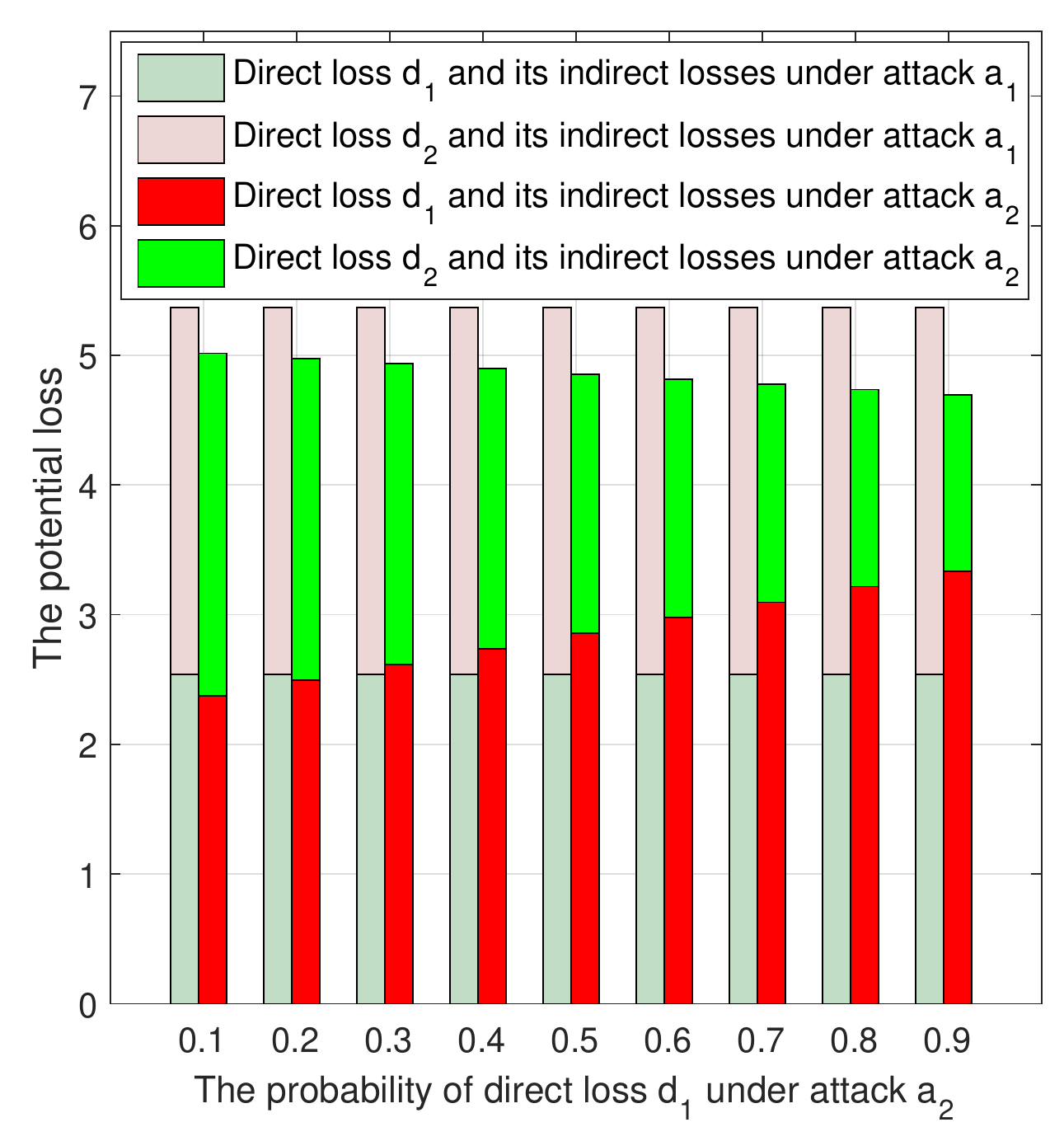} &
		\epsfxsize=1.65 in \epsffile{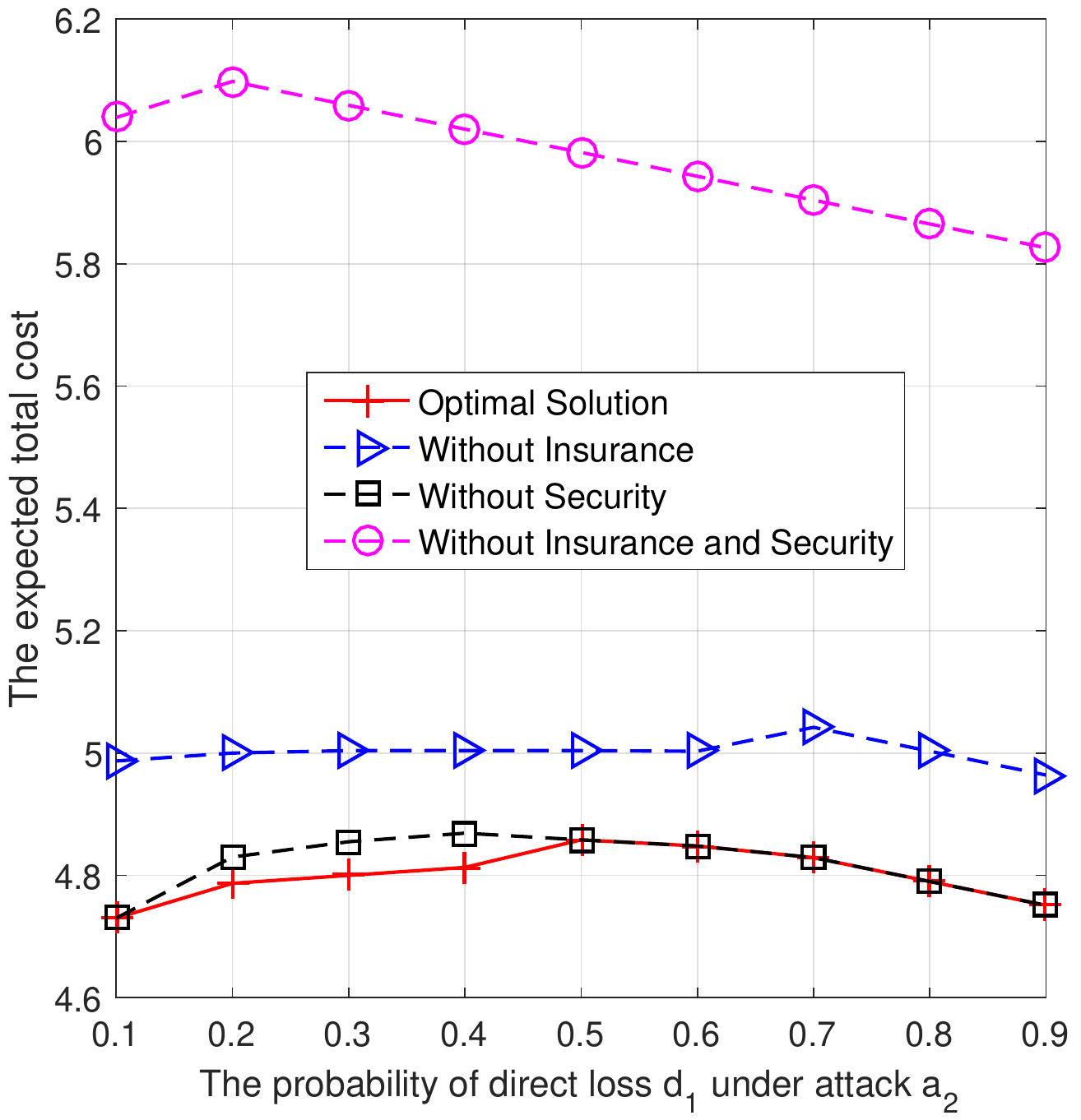} \\ [-0.2cm]
		(a) & (b) 
		\end{array}$
		\caption{The overall network throughput of the secondary system in the RFPB-CRN.}
		\label{fig:vary_d2_a2}
	\end{center}
\end{figure}

We then set the probability of attack $a_1$ at $0.4$ and vary the probability of direct loss $d_1$ of attack $a_2$ to evaluate its impacts to the security policy of the CSP. In Table~\ref{tab:Policy2}, when the probability of direct loss $d_1$ under attack $a_2$, i.e., $p^{d_1}_{a_2}$, is lower than $0.5$, the CSP will invest in security and insurance packages. However, when $p^{d_1}_{a_2}$ is greater than $0.5$, the CSP only spends money for repair packages to prevent the potential indirect loss of direct loss $d_1$ under attack $a_2$. The reason can be explained through Fig.~\ref{fig:vary_d2_a2}(a). In particular, in Fig.~\ref{fig:vary_d2_a2}(a), as $p^{d_1}_{a_2}$ increases, the potential loss of direct loss $d_1$ together with its indirect loss under attack $a_2$ will be increased. However, the total potential loss of attack $a_2$ is slightly reduced because under attack $a_2$ the direct loss $d_1$ is lower than direct loss $d_2$. Thus, instead of using security and/or insurance packages which have higher costs, the CSP will choose only a repair package to deal with direct loss $d_2$ under attack $a_2$ when $p^{d_1}_{a_2}$ is high. In Fig.~\ref{fig:vary_d2_a2}(b), we evaluate the performance of the proposed solution by comparing its expected total cost with those of other approaches. As $p^{d_1}_{a_2}$ increases from $0.1$ to $0.5$, the expected total cost of the CSP slightly increases. However, when $p^{d_1}_{a_2}$ keeps increasing from $0.5$ to $0.9$, the expected total cost of the CSP slightly decreases. Again, the expected total cost obtained by the proposed solution is always lower than those of other approaches.

%++++++++++++++++++++++++++++++++++++++++++++++++++++++++++++++++++++
%++++++++++++++++++++++++++++++++++++++++++++++++++++++++++++++++++++
\section{Summary} 
\label{sec:Sum}

In this paper, we have developed a new framework based on stochastic programming approach for the risk management problem in mobile cloud environment. The aim of this framework is to find the optimal tradeoff among security policies, insurance policies, and countermeasures under uncertainty of cyberattacks and their losses such that the expected total cost of the cloud service provider is minimized. Through numerical results, we have shown the effectiveness as well as flexibility of the proposed solution in dealing with cyberattacks. The findings in this paper are especially important not only for cloud service providers in implementing security policies, but also for security and insurance providers in proposing appropriate offers to attract more customers. For the future work, we will study the relation between security and insurance providers through bundling strategies and matching theory. Furthermore, the relation between a direct loss and its indirect losses will be further investigated.

%%++++++++++++++++++++++++++++++++++++++++++++++++++++++++++++++++++++
%%++++++++++++++++++++++++++++++++++++++++++++++++++++++++++++++++++++
%\appendices
%
%%+++++++++++++++++++++++++++++
%\section{The proof of Theorem~\ref{theo:opt_data_trans_period}}
%\label{app:theo:opt_data_trans_period}

%++++++++++++++++++++++++++++++++++++++++++++++++++++++++++++++++++++
%++++++++++++++++++++++++++++++++++++++++++++++++++++++++++++++++++++
\bibliographystyle{IEEE}

\end{document}